\documentclass[aps,prl,reprint,superscriptaddress]{revtex4-2}
\pdfoutput=1 %For arXiv.org - usually not necessary
\usepackage{amsmath,amssymb,amsthm}
\usepackage{wasysym}
\usepackage{graphicx, color}
\usepackage[english]{babel}

\usepackage{hyperref}
\hypersetup{colorlinks=true,linkcolor=black,urlcolor=blue,citecolor=blue,
  pdfinfo={
    Title={Global stability of fluid flows despite transient growth of energy},
    Author={Federico Fuentes, David Goluskin, Sergei Chernyshenko},
    Subject={Stability in fluid dynamics},
    Keywords={laminar flow, fluid dynamics, Lyapunov stability, energy stability, sum-of-squares polynomials, polynomial optimization}
  }
}

\definecolor{mygreen}{rgb}{0,.5,0}
\definecolor{mybrown}{rgb}{.43,.21,.1}

\newcommand{\beq}{\begin{equation}}
\newcommand{\eeq}{\end{equation}}

\newcommand{\ab}{\mathbf{a}}
\newcommand{\ff}{\mathbf{f}}
\newcommand{\hh}{\mathbf{h}}
\newcommand{\nn}{\mathbf{n}}
\newcommand{\xx}{\mathbf{x}}
\newcommand{\uu}{\mathbf{u}}
\newcommand{\vv}{\mathbf{v}}
\newcommand{\ww}{\mathbf{w}}
\newcommand{\BB}{\mathsf{B}}
\newcommand{\CC}{\mathbf{c}}
\newcommand{\DD}{\mathsf{D}}
\newcommand{\UU}{\mathbf{U}}
\newcommand{\I}{\mathrm{i}}
\newcommand{\cL}{\mathcal{L}}
\newcommand{\ddt}{\frac{\rm d}{{\rm d}t}}
\newcommand{\dx}{{\rm d}\xx}

\newcommand{\Rey}{\mathrm{Re}}

\newcommand{\Thetab}{\boldsymbol{\Theta}}

\newcommand{\Thetabab}{\boldsymbol{\Theta}_{\mathbf{AB}}}
\newcommand{\Thetabc}{\boldsymbol{\Theta}_{\mathbf{C}}}

\newcommand{\Thetaabi}{\Theta_{\mathbf{AB}i}}
\newcommand{\Thetaci}{\Theta_{\mathbf{C}i}}
\newcommand{\Mb}{\mathbf{M}}
\newcommand{\tder}[2]{\frac{\mathrm{d} #1}{\mathrm{d} #2}}
\newcommand{\pder}[2]{\frac{\partial #1}{\partial #2}}
\newcommand{\psderm}[3]{\frac{\partial^2 #1}{\partial #2 \partial #3}}
\newcommand{\scalar}[2]{\langle #1\, ,#2\rangle}
\newcommand{\T}{\mathsf{T}}
\newcommand{\eps}{\varepsilon}

\newcommand{\GG}{\mathsf{G}}
\newcommand{\R}{\mathbb{R}}
\newcommand{\N}{\mathbb{N}}
\newcommand{\Real}{\mathfrak{Re}}

\begin{abstract}
Verifying nonlinear stability of a laminar fluid flow against all perturbations is a central challenge in fluid dynamics. Past results rely on monotonic decrease of a perturbation energy or a similar quadratic generalized energy. None show stability for the many flows that seem to be stable despite these energies growing transiently. Here a broadly applicable method to verify global stability of such flows is presented. It uses polynomial optimization computations to construct non-quadratic Lyapunov functions that decrease monotonically. The method is used to verify global stability of 2D plane Couette flow at Reynolds numbers above the energy stability threshold found by Orr in 1907. This is the first global stability result for any flow that surpasses the energy method.
\end{abstract}

\begin{document}

\title{Global stability of fluid flows despite transient growth of energy}
\author{Federico Fuentes}
\affiliation{Department of Mathematics, Cornell University, Ithaca, NY 14853, USA}
\affiliation{Instituto de Ingenier\'ia Matem\'atica y Computacional, Pontificia Universidad Cat\'olica de Chile, Macul, Santiago 7820436, Chile}
\author{David Goluskin}
\affiliation{Department of Mathematics and Statistics, University of Victoria, Victoria, BC, V8P 5C2, Canada}
\author{Sergei Chernyshenko}
\affiliation{Department of Aeronautics, Imperial College London, London SW7 2AZ, UK}

\maketitle

A central approach to understanding fluid dynamics has been to study a handful of canonical systems in detail. Despite many discoveries over the last century, one of the simplest-seeming questions remains open for some of the most-studied systems: at given parameter values, will the flow return to its simplest (laminar) state no matter how it is perturbed? Laboratory experiments and simulations of the Navier--Stokes equations are unable to give a complete answer for all perturbations. Theoretical methods are needed to guarantee global stability.

For a steady laminar velocity field $\UU(\xx)$ solving the incompressible Navier--Stokes equations, the velocity, $\uu(\xx,t)$, and pressure, $p(\xx,t)$, of perturbations around the laminar state evolve according to
\begin{align}
\textstyle{\pder{}{t}}\uu+\uu\cdot\nabla\uu&=-\nabla p+\tfrac{1}{\Rey}\Delta\uu + A(\uu), \label{eq:NS} \\
\nabla\cdot\uu &= 0, \label{eq:inc}
\end{align}
where $A(\uu)= -\UU\cdot\nabla\uu - \uu\cdot\nabla\UU$ and $\Delta$ is the Laplacian operator~\cite{serrin1959}. Quantities in \eqref{eq:NS}--\eqref{eq:inc} are dimensionless, having been scaled using a length scale $h$, velocity scale $U$, and kinematic viscosity $\nu$. Choices of $h$ and $U$ depend on the particular system. The dimensionless Reynolds number is $\Rey=Uh/\nu$.

There is a critical threshold $\Rey_G>0$ such that the laminar state $\UU$ is globally asymptotically stable (meaning all perturbations $\uu$ eventually converge to zero) if and only if $\Rey<\Rey_G$ \cite{serrin1959}. Loss of global stability is not sufficient for turbulence, but it is necessary, and often it is more informative than linear stability. Linear stability of the laminar state does not preclude turbulence whose onset is subcritical \cite{daviaud1992,romanov1973,willis2008,sano2016,carlson1982flow}, nor does it ensure that the laminar state is physically realizable because the basin of attraction can be minuscule \cite{chapman2002,grossmann2000,bedrossian2017}. The value of $\Rey_G$, however, can be very hard to determine.

An upper bound on $\Rey_G$ is provided by any $\Rey$ at which a sustained non-laminar flow is found. A lower bound on $\Rey_G$ requires finding a $\Rey$ threshold below which the laminar state is globally stable. Thus far the only method applicable to all systems governed by \eqref{eq:NS}--\eqref{eq:inc} has been the energy method pioneered by Reynolds and Orr \cite{reynolds1895,orr}, where one finds the threshold $\Rey_E$ such that the kinetic energy, $E=\frac{1}{2}\int |\uu|^2\dx$, of every perturbation decreases monotonically toward zero if and only if $\Rey<\Rey_E$. Often the lower bound on $\Rey_G$ provided by $\Rey_E$ is very conservative. In systems where turbulence is driven by parallel shear, such as pressure-driven flow in a pipe or boundary-driven flow in a layer, the energy stability thresholds $\Rey_E$ \cite{joseph1966,joseph1969,busse1972,josephbook} are much smaller than the minimum $\Rey$ at which sustained non-laminar states have been found \cite{zahn1974,nagata,waleffe2003,willis2008}. In other words, there is a large gap between these lower and upper bounds on $\Rey_G$.

Global stability at $\Rey$ values above $\Rey_E$ has been shown only in special cases where the energy method can be slightly generalized. Each such result has relied on monotonic decrease of a quadratic integral that is an inviscid invariant, meaning the nonlinear term in \eqref{eq:NS} does not contribute to the expression for the integral's evolution. For symmetric perturbations where individual components of $E$ are conserved, for instance, one can consider various linear combinations of these components \cite{josephbook,straughan2013,vonwahl2005,Galdi1990}. Lacking an artificial symmetry on $\uu$, however, $E$ is the only nonnegative quadratic integral that can be shown to decrease globally. In this general situation there has been no method for verifying global stability above $\Rey_E$, aside from the one presented here.

The standard way to show that a solution of a dynamical system is globally asymptotically stable is to construct a Lyapunov function. Here this is a functional $V$ that maps each spatial function $\uu(\cdot,t)$ to a real number and satisfies $V(\mathbf 0)=0$. Let $\cL V$ denote the Lie derivative of $V$ along PDE solutions of \eqref{eq:NS}--\eqref{eq:inc}, meaning $\cL V$ is the functional such that $\cL V(\uu(\cdot,t)) =\ddt V(\uu(\cdot,t))$ for all $\uu(\xx,t)$ solving \eqref{eq:NS}--\eqref{eq:inc}. The $\uu=\mathbf 0$ state is globally attracting if $V(\uu)>0$ and $\cL V(\uu)<0$ for all nonzero $\uu$ admitted by the boundary conditions \cite{mironchenko2019}. The energy method uses $V=E$ or, when symmetries allow it, weighted combinations of the components of $E$.

Our method constructs Lyapunov functionals $V$ with polynomial dependence on $\uu$, in particular with
\beq
V(\uu)=V(\ab,q)=E^d+ P(\ab,q),
\label{eq:Vansatz}
\eeq
where $\ab(\uu)\in\R^m$, $q(\uu)\in\R$, $d$ is an integer, and $P$ is a polynomial whose degree is at most $2d-1$. By definition, the components of $\ab$ are projections of $\uu$ onto an orthogonal set of spatial modes, $\{\uu_1(\xx),\ldots,\uu_m(\xx)\}$, and $\tfrac{1}{2}q^2$ is the energy of the unprojected remainder of $\uu$. For reasons explained shortly, we choose the $\uu_i$ to be eigenfunctions of the energy stability operator. Constructing $P$ and verifying that $V$ is a valid Lyapunov functional presents major challenges beyond the quadratic case. A general way to surmount these challenges is presented below, but first we summarize stability results found by applying our method to a classic fluid flow.

To show that Lyapunov functionals of the form \eqref{eq:Vansatz} can surpass existing methods we consider 2D plane Couette flow, which is driven by parallel relative motion of the boundaries. We have verified global stability of this flow beyond the energy stability threshold given by Orr in 1907 \cite{orr}. The reason for considering a 2D flow, aside from Orr's result being especially longstanding, is to reduce the computational cost of testing our method. The same approach is applicable to arbitrary 3D perturbations, but this is left for future work. The flow is periodic in the streamwise direction, $x\in(0,L)$, and confined in the wall-normal direction, $y\in(-\frac{1}{2},\frac{1}{2})$. Perturbations about the laminar flow $\UU=(y,0)$ obey \eqref{eq:NS}--\eqref{eq:inc} and satisfy no-slip conditions $\uu(x,\pm\frac{1}{2})=\mathbf{0}$ at the walls. With this nondimensionalization, $\Rey$ is defined using the full velocity difference and height difference between the shearing planes. Some authors use half these differences, so their Reynolds number is $1/4$ of the $\Rey$ shown here.

The true value of $\Rey_G$ in 2D plane Couette flow is unknown. Several computational efforts have failed to find sustained non-laminar states \cite{orszag1980, rincon2007, ehrenstein2008}, and the laminar state is linearly stable for all $\Rey$ \cite{romanov1973}, so $\Rey_G$ has no known upper bound and may be infinite. For each streamwise period $L$, the energy method gives a lower bound $\Rey_E(L)\le\Rey_G(L)$. As found by Orr \cite{orr}, its minimum $\Rey_E\approx177.2$ occurs at integer multiples of $L_E\approx1.659$. (In 3D, $\Rey_G$ is bounded below by $\Rey_E\approx82.6$ \cite{joseph1966, busse1972} and above by 511, the smallest $\Rey$ at which traveling waves solutions have been computed numerically \cite{nagata,waleffe2003}.)

\begin{figure}[t]
\centering 
\includegraphics[width=8cm,trim=0 0 0 10]{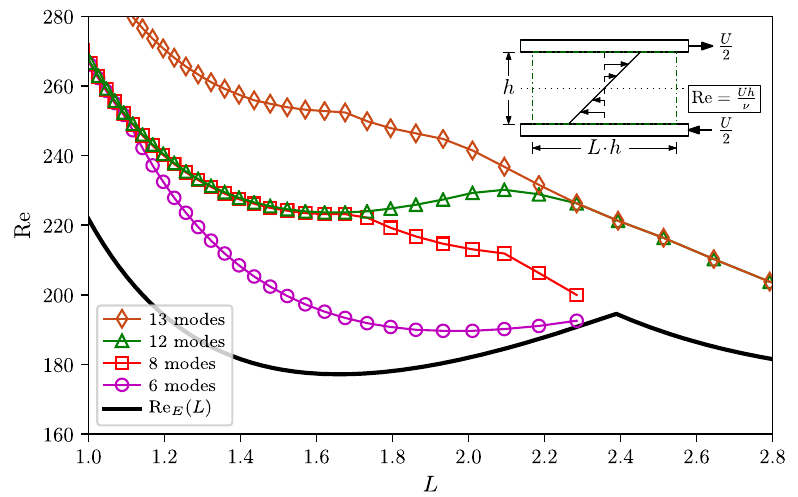} 
\vspace{-3.5mm}
\caption{\label{fig:GS2DPlaneCouette}Reynolds numbers ($\Rey$) at which laminar plane Couette flow is globally asymptotically stable against 2D perturbations of period $L$. Each symbol indicates values $(\Rey,L)$ where we verified stability using a quartic Lyapunov functional. Each functional depends explicitly on the flow's projection onto 6, 8, 12, or 13 energy eigenmodes, and on the unprojected energy. Lines connect symbols to guide the eye. Orr's energy stability threshold $\Rey_E(L)$ is also shown.}
\end{figure}

Here we have constructed many $V$ of the form \eqref{eq:Vansatz}, all having quartic degree ($d=2$) and depending explicitly on the projections $a_i$ of $\uu$ onto various $\uu_i$ modes. Results are reported for four different mode sets (defined later) whose number of modes ($m$) are 6, 8, 12, and 13. Figure~\ref{fig:GS2DPlaneCouette} shows $\Rey$ values at which stability has been verified using each set of modes, along with the energy stability threshold $\Rey_E(L)$. At each plotted point, a different Lyapunov functional was constructed to show global stability for perturbations of period $L$ at the $\Rey$ indicated. Raising the number of modes on which $V$ depends increases the $\Rey$ at which stability can be verified, but it also increases the computational cost of constructing $V$ by the method explained below, which limited us to 13 modes.

Over the full range of periods $L$ for which computations were performed, results surpass the energy method. For instance, at the most energy-unstable period $L_E$ where the energy method gives stability up to $\Rey_E\approx177.2$, our best $V$ verified stability at $\Rey=252.4$. Beyond the implications for Couette flow, the greater significance of these results is the proof of concept for a broadly applicable new method---the first generalization of the energy method that is applicable to any 2D or 3D flow.

To recall the workings of the energy method, note that positivity of $E$ is clear, so implementing the energy method amounts to determining the $\Rey$ at which $\cL E<0$ for all admissible perturbations. In systems where $\uu$ is periodic and/or vanishes at all boundaries,
\beq
\cL E = \int\left(-\tfrac{1}{\Rey}|\nabla\uu|^2 - \uu\cdot\DD\cdot\uu\right)\dx,
\eeq
where $\DD=\tfrac{1}{2}(\nabla\UU+\nabla^\T\UU)$ is the laminar strain-rate tensor \cite{serrin1959}.
Variational arguments imply that $\cL E<0$ for all divergence-free nonzero $\uu$ if and only if all eigenvalues $\lambda$ are negative for the energy eigenproblem \cite{josephbook,drazin2004hydrodynamic,doering1995}
\beq
	\left(\tfrac{1}{\Rey}\Delta-\DD\right)\ww -\nabla\zeta =\lambda\ww, \quad\,\, \nabla\cdot\ww=0,
	\label{eq:eigen}
\eeq 
where $\zeta$ is the Lagrange multiplier enforcing incompressibility of $\ww$. The largest $\Rey$ at which $\lambda\le0$ defines the energy stability threshold $\Rey_E$. Only because $\cL E$ is quadratic can its negativity be verified from a linear Euler--Lagrange equation \eqref{eq:eigen}. Going beyond quadratic $V$ requires another way to enforce $V>0$ and $\cL V<0$.

To construct new non-quadratic $V$, we follow the ideas in \cite{ChernyshenkoGoulart} and consider a partial Galerkin expansion of $\uu$,
\beq
	\uu(\xx,t)=\sum_{i=1}^m a_i(t)\uu_i(\xx)+\vv(\xx,t)\,,
\label{eq:Galerkin}
\eeq
where the $\uu_i$ are selected modes of the energy eigenproblem \eqref{eq:eigen}, and $a_i=\int\uu\cdot\uu_i\,\dx$ is the orthogonal projection of $\uu$ onto $\uu_i$. Let $q=(\int|\vv|^2\dx)^{1/2}$, so the perturbation energy is $E=\tfrac{1}{2}(|\ab|^2+q^2)$. Lyapunov functionals $V$ will be functions of the $m+1$ scalars $(\ab,q)$, each of which is a functional of $\uu$.

To derive the functional $\cL V$ that coincides with $\tder{}{t}V$ along solutions of \eqref{eq:NS}--\eqref{eq:inc}, we let only even powers of $q$ appear in $V$, in which case $\tder{}{t}V=\pder{V}{\ab}\cdot\tder{\ab}{t}+\pder{V}{q^2}\tder{q^2}{t}$. Projecting the Navier--Stokes equations gives expressions of the form $\tder{\ab}{t}=\ff+\Thetab$ and $\tder{q^2}{t}=-2\ab\cdot\Thetab+2\Gamma$ \cite{huang2015}. These constitute an ``uncertain system'' for the evolution of $(\ab,q)$ since $\Thetab$ and $\Gamma$ (given below) depend on the tail $\vv$ in a way that is not uniquely determined by its energy $\tfrac{1}{2}q^2$. The resulting expression for $\cL V$ is \cite{huang2015}
\beq
	\begin{aligned}
	\cL V(\ab,q,\vv) &=G(\ab,q,\vv)+\Mb(\ab,q)\cdot\Thetab(\ab,\vv), ~\text{with}\\
	G(\ab,q,\vv)&=\pder{V}{\ab}\cdot\ff(\ab)+2\pder{V}{q^2}\Gamma(\vv),\\
	\Mb(\ab,q)&=\pder{V}{\ab}-2\pder{V}{q^2}\ab,\\
	\Thetab(\ab,\vv)&=\Thetabab(\ab,\vv)+\Thetabc(\vv), \\
	f_i(\ab)&=L_{ij}a_j+N_{ijk}a_ja_k, \\
	L_{ij}&=\tfrac{1}{\Rey}\scalar{\uu_i}{\Delta\uu_j}+\scalar{\uu_i}{A(\uu_j)}, \\
	N_{ijk}&=-\scalar{\uu_i}{\uu_j\cdot\nabla\uu_k}, \\
	\Thetaabi(\ab,\vv)&=\scalar{\vv}{\hh_{i0}}+\scalar{\vv}{\hh_{ij}}a_j, \\
	\hh_{i0}&=\tfrac{1}{\Rey}\Delta\uu_i+\UU\cdot\nabla\uu_i-\uu_i\cdot\nabla^\T\UU, \\
	\hh_{ij}&=\uu_j\cdot\nabla\uu_i-\uu_i\cdot\nabla^\T\uu_j, \\
	\Thetaci(\vv)&=\scalar{\vv}{\vv\cdot\nabla\uu_i}, \\
	\Gamma(\vv)&=\tfrac{1}{\Rey}\scalar{\vv}{\Delta\vv}-\scalar{\vv}{\DD\vv},
	\end{aligned}
\label{eq:auxiliary}
\eeq
and $\scalar{\uu}{\vv}=\int\uu\cdot\vv\,\dx$.

Positivity of $V$ is enforced by regarding $V(\ab,q)$ as a polynomial on $\mathbb R^{m+1}$, rather than a functional of $\uu$. Requiring positivity of this polynomial away from the origin constrains $P$. Negativity of $\cL V$ is enforced in a similar way, but since $\cL V$ depends on the full tail $\vv$, it first must be bounded above by a polynomial depending only on $(\ab,q)$. The reason we choose the $\uu_i$ to be modes of the energy eigenproblem is so that $\Gamma(\vv)\leq\kappa q^2$~\cite{ChernyshenkoGoulart}, where $\kappa$ is the largest eigenvalue from \eqref{eq:eigen} not associated with any of the $m$ modes in the sum of \eqref{eq:Galerkin}. Enough modes are included so that $\kappa<0$, and we impose $\pder{V}{q^2}\geq0$ so that
\beq
G(\ab,q,\vv)\le\tilde{G}(\ab,q)=\pder{V}{\ab}\cdot\ff(\ab)+2\pder{V}{q^2}\kappa q^2.
\label{eq:G}
\eeq
A procedure described in the Supplement introduces a polynomial $\Xi(\ab,q)$ with auxiliary constraints that ensure
\beq
\Mb(\ab,q)\cdot\Thetab(\ab,\vv)\le\Xi(\ab,q).
\label{eq:Xi}
\eeq
By \eqref{eq:auxiliary}--\eqref{eq:Xi}, if $\tilde G+\Xi<0$ for all $(\ab,q)$, then $\cL V<0$ for all $\uu$. Therefore, if polynomials $P(\ab,q)$ and $\Xi(\ab,q)$ are found such that $V>0$, $\tilde{G}+\Xi<0$, and $\pder{V}{q^2}\geq0$ for all nonzero $(\ab,q)$, and such that the inequalities in the Supplement guaranteeing \eqref{eq:Xi} hold, then $V$ is a valid Lyapunov functional. Each of these constraints amounts to nonnegativity of a polynomial expression.

Verifying that a polynomial is nonnegative is computationally intractable (NP-hard) in general \cite{Murty1987}. A tractable sufficient condition is that the polynomial can be written as a sum of squares of other polynomials. Computational techniques for enforcing sum-of-squares (SOS) constraints, introduced two decades ago \cite{Nesterov2000, ParriloThesis, Lasserre2001}, let us search for $P$ and $\Xi$ in a chosen bounded-degree set of polynomials subject to SOS constraints that imply all of the inequalities described above. If such $P$ and $\Xi$ are found, then $V$ defined by \eqref{eq:Vansatz} is a valid Lyapunov functional. The tunable coefficients of $P$ and $\Xi$ appear linearly in the expressions that must be SOS, and the problem of choosing these coefficients subject to the SOS constraints can be reformulated \cite{Parrilo2013a, lasserre2015} as a semidefinite program---a type of conic optimization problem that can be solved numerically using specialized software. 
When $\Rey<\Rey_E$, a solution always exists with $P=\Xi=0$.

The approach to fluid stability described above was proposed but not implemented in \cite{ChernyshenkoGoulart}. As a preliminary test, the idea was applied in \cite{huang2015} to an example contrived to have simple energy eigenmodes. Quartic and sextic Lyapunov functionals were successfully computed in \cite{huang2015}, but they had no chance to improve upon the energy method; a weighted energy (which can be used due to symmetries) already gives $\Rey_G$ exactly for that flow. The present work adds three contributions. First, we show that the approach of \cite{ChernyshenkoGoulart} can surpass quadratic Lyapunov functionals in practice. Second, we do this in a realistic context where the energy eigenproblem \eqref{eq:eigen} must be solved computationally. Third, we make a crucial technical change to the way $\Xi$ is defined and constrained in  \cite{ChernyshenkoGoulart}, as described in the Supplement, and this improves our results dramatically.

The ansatz \eqref{eq:Vansatz} for $V$ is not an arbitrary polynomial since some structure can be deduced \emph{a priori}. Both $V$ and $\cL V$ must be sign-definite, so their highest-degree terms must be of even degree. This is possible only if the nonlinearity in \eqref{eq:NS} does not contribute to the evolution of the highest-degree term in $V$, in which case both expressions can have the same maximum degree. This is why the leading term in \eqref{eq:Vansatz} takes the form $E^d$. Further, $P$ can have no terms of degree less than two since $V$ must have a unique minimum when $\uu=\mathbf 0$. When $d=1$ these constraints require $V$ to be the energy $E$ in general, reflecting the lack of freedom in the quadratic case. When $d\ge2$ there is significant freedom in the choice of $P$.

Constructing a polynomial $\Xi$ that is guaranteed to satisfy~\eqref{eq:Xi} requires computing all tensors in \eqref{eq:auxiliary}. To do so one must first compute energy eigenmodes of \eqref{eq:eigen} for the chosen values of $(\Rey,L)$ and then select the set of modes $\{\uu_1,\ldots,\uu_m\}$, where $V$ will depend explicitly on projections of $\uu$ onto these modes. It is necessary to include all modes with positive eigenvalues at the given $(\Rey,L)$, so that $\kappa<0$ in \eqref{eq:G}, and to include enough stable modes that trajectories of the truncated system $\tder{\ab}{t}=\ff$ are bounded. Beyond this, there is freedom in the number and choice of modes. For a fixed number of modes, experimentation may be needed to determine which mode set gives the strongest stability results.

\begin{figure}[t]
\centering 
\includegraphics[width=8cm]{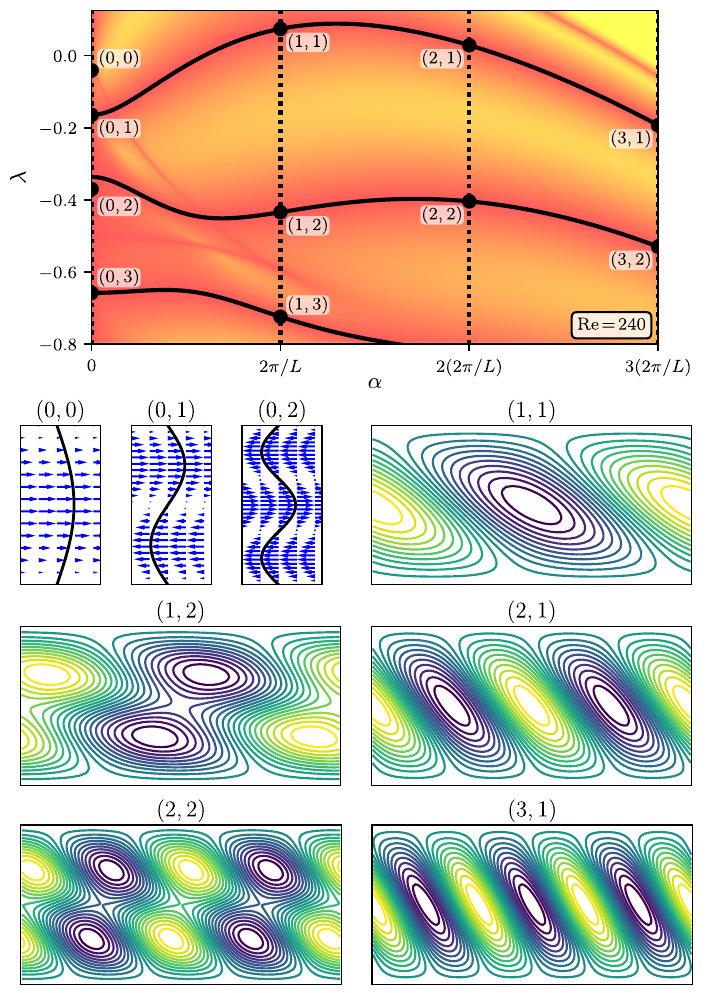} 
\vspace{-1mm}
\caption{Energy stability eigenmodes for $(\Rey,L)=(240,2)$. The top panel shows eigenvalues as a function of streamwise wavenumber $\alpha$. Shading indicates the minimum singular value of a boundary constraint matrix $\BB$ (cf.\ the Supplement); black curves are zeros of this minimum, corresponding to eigenvalues. Eigenmodes consistent with $L=2$ occur at multiples of $\frac{2\pi}{L}$, marked by black dots. The mode with the $j^{th}$ largest eigenvalue among eigenmodes with wavenumber $\alpha=\frac{2\pi i}{L}$ is labeled $(i,j)$. Bottom panels show streamlines for selected modes. When $i\neq0$ we include two modes, shifted by a quarter-period in $x$, to span the relevant eigenspace. The 6-mode set consists of the $(0,0)$, $(0,1)$, $(1,1)$, and $(1,2)$ eigenmodes. The 8-mode set adds $(2,1)$, the 12-mode set adds $(2,2)$ and $(3,1)$, and the 13-mode set adds $(0,2)$.
\label{fig:Eigenmodes}}
\end{figure}

To apply our method to 2D plane Couette flow, we first solve the energy eigenproblem \eqref{eq:eigen} as detailed in the Supplement. The eigenproblem must be solved anew for each $L$ and $\Rey$ considered, giving eigenfunctions whose streamwise wavenumbers $\alpha$ are multiples of $\tfrac{2\pi}{L}$. As an example, Fig.~\ref{fig:Eigenmodes} shows eigenvalues and corresponding eigenmodes for $(\Rey,L)=(240,2)$, a point in the parameter regime where energy can grow transiently yet our computations verify stability.

The four nested sets of eigenmodes $\{\uu_1,\ldots,\uu_m\}$ that were used to compute the stability results of Fig.~\ref{fig:GS2DPlaneCouette} are defined in the Fig.~\ref{fig:Eigenmodes} caption. For each $(\Rey,L)$ and set of modes, all tensors in \eqref{eq:auxiliary} were computed numerically. We then formulated the SOS computations described above, searching for polynomials $P$ and $\Xi$ such that $V$ was verified to be a Lyapunov functional. The parser YALMIP \cite{Lofberg2004,Lofberg2009} was used to reformulate all SOS constraints as semidefinite programs, which were then solved using MOSEK \cite{mosek}. The resulting $P$ and $\Xi$ have many terms, so we do not report them here.

For each $L$ and set of modes, the symbol plotted in Fig.~\ref{fig:GS2DPlaneCouette} is the largest $\Rey$ for which our SOS computations found a valid quartic $V$. We expect the stability thresholds in Fig.~\ref{fig:GS2DPlaneCouette} will continue to improve with an increase to the number of eigenmodes ($m$) on which $V$ explicitly depends in \eqref{eq:Vansatz}. However, our computations for 13 modes are already expensive. This prevents us from considering very large $L$ since the number of modes that would be needed grows at least linearly with $L$. Thus the present version of our method cannot apply to arbitrary-$L$ perturbations in very long domains, although it surpasses the energy method for perturbations up to whatever period is computationally tractable. 
Aside from adding modes, stability thresholds could be improved by raising the polynomial degree of $V$, but sextic $V$ demand much larger computational cost and memory footprint.

As an independent check that the $V$ constructed by our SOS computations decrease monotonically in time, we numerically integrated \eqref{eq:NS}--\eqref{eq:inc} for 2D Couette flow using the code Dedalus \cite{dedalus}, starting from $10^4$ random initial conditions (cf.\ the Supplement) in the energy-unstable case $(\Rey,L)=(240,2)$. In all simulations our $V$ depending on 13 modes decreased monotonically, whereas $E$ increased transiently in 7 simulations.

In summary, we have presented a general method for constructing polynomial Lyapunov functionals to show global stability of fluid flows. It may be used to surpass the many conservative results derived using energy (or other quadratic integrals) to which past studies of fluid stability have been confined. Our approach is more technical than the energy method but can be implemented using modern computational tools of polynomial optimization. We have verified stability for 2D plane Couette in a regime where energy grows transiently. This improves on a century-old stability criterion of Orr, at least for perturbations whose streamwise periods are not too large. As far as we know, this is the first global stability result for any flow that is stronger than what can be shown using the energy method or its generalizations to other quadratic integrals. The natural next step is to apply the same approach to 3D perturbations of plane Couette flow or another 3D flow where the energy method is overly conservative, such as pipe flow. The procedure will be the same as in the present 2D example, only with greater technicality and computational cost. 

\begin{acknowledgments}
The authors thank the Geophysical Fluid Dynamics program at Woods Hole Oceanographic Institution, which hosted two of us (FF and DG) during much of this work. Many helpful discussions with Giovanni Fantuzzi are appreciated, and one of us (FF) thanks A.~Townsend for computer resources provided at Cornell University. During this work, FF was supported by the National Science Foundation (NSF) award 2012658, DG was supported by the NSERC Discovery Grants Program through awards RGPIN-2018-04263, RGPAS-2018-522657, and DGECR-2018-00371, and SC was supported by the UK Engineering and Physical Sciences Research Council grant EP/J011126/1.
\end{acknowledgments}

%\bibliographystyle{apsrev4-2}
%\bibliography{main}

%\end{document}

\bibliographystyle{apsrev4-2}
\bibliography{main}

\newpage {\hspace{2mm}}
\newpage
\setcounter{section}{1}
\setcounter{equation}{0}
\renewcommand{\theequation}{S\arabic{equation}}

\onecolumngrid
\begin{center}
	{\large\hspace*{20mm}\textbf{Supplementary Material:}}
	\vspace{1mm}
	\newline{\hspace*{10mm}\large\textbf{Global stability of fluid flows despite transient growth of energy}}
	\vspace{3mm}
	\newline{\normalsize Federico Fuentes, David Goluskin, Sergei Chernyshenko}
	\vspace{4mm}
\end{center}
\twocolumngrid

\subsection{Lyapunov stability theorem}

To show that the energy of the velocity perturbations $\uu(\xx,t)$ satisfying \eqref{eq:NS}--\eqref{eq:inc} tends to zero as $t\to\infty$, we use the Lyapunov stability theorem given as Proposition 3.2 in \cite{mironchenko2019}. Results in that work apply to the present infinite-dimensional dynamics, whereas most Lyapunov stability theorems are stated for ordinary differential equations. Let $X$ be the class of differentiable spatial functions satisfying the no-slip and periodic boundary conditions imposed by the fluid domain. Consider $X$ as a normed linear space with the spatial $L^2$ norm, $\|\uu\|^2=\int|\uu|^2\,\dx$. A Lyapunov stability result in this setting applies to every solution of \eqref{eq:NS}--\eqref{eq:inc} that remains in the class $X$ for $t\geq0$. If a Lyapunov functional $V:X\to\R$ can be constructed satisfying suitable conditions, then $\lim_{t\to\infty}\|\uu(\cdot,t)\|=0$ for every admissible initial condition $\uu(\cdot,0)\in X$. In other words, the laminar state $\uu=\mathbf{0}$ would be globally asymptotically stable.

The conditions in Proposition 3.2 of \cite{mironchenko2019} require that $V$ is continuous and satisfies
\begin{gather}
	\xi_L(\|\uu\|)\leq V(\uu)\leq\xi_U(\|\uu\|), \label{eq:VCond1} \\
	\cL V(\uu)\leq-\xi_D(\|\uu\|), \label{eq:VCond2}
\end{gather}
where $\xi_L$, $\xi_U$ and $\xi_D$ are continuous strictly increasing functions from $\R_{+}$ to $\R_{+}$ that vanish at $0$, and where $\xi_L$ and $\xi_U$ are unbounded. The functional $\cL V:X\to\R$ is the Lie derivative of $V$ along solutions of \eqref{eq:NS}--\eqref{eq:inc}. It has no explicit time-dependence but is defined such that $\cL V(\uu(\cdot,0))=\ddt V(\uu(\cdot,t))\big|_{t=0}$ for the solution $\uu(\xx,t)$ of the Navier--Stokes equations \eqref{eq:NS}--\eqref{eq:inc} with initial condition $\uu(\cdot,0)\in X$.

Condition \eqref{eq:VCond1} implies that $V(\mathbf{0})=0$, and the fact that the laminar flow $\uu=\mathbf{0}$ is an equilibrium point implies $\cL V(\mathbf{0})=0$. Conditions \eqref{eq:VCond1} and \eqref{eq:VCond2} require that $V(\uu)>0$ and $\cL V(\uu)<0$ for all nonzero $\uu$, and that $V$ approaches infinity as $\|\uu\|$ does (i.e., $V$ is radially unbounded). In the present work we impose constraints which ensure $V(\uu)\geq\eps E(\uu)=\eps\tfrac{1}{2}\|\uu\|^2=\xi_L(\|\uu\|)$ for some $\eps>0$. Since $V$ is a polynomial of the form \eqref{eq:Vansatz}, this implies $V\leq c_aE+c_bE^d=\xi_U$ for some positive constants $c_a$ and $c_b$, so condition \eqref{eq:VCond1} holds. Our constraints also ensure that there exists a function $Q(\ab,q)$ such that $\cL V(\uu)\leq Q(\ab,q)\leq-\eps\tfrac{1}{2}\|\uu\|^2$ for some $\eps>0$, so condition \eqref{eq:VCond2} holds with $\xi_D=\eps E$. Thus, Proposition 3.2 of \cite{mironchenko2019} can be applied.   

\subsection{Constraints on \texorpdfstring{$\Xi$}{Xi}}

The function $Q(\ab,q)$ that must bound $\cL V$ above takes the form $Q=\tilde{G}+\Xi$, where $\tilde{G}$ is defined by \eqref{eq:G} and $\Xi$ satisfies \eqref{eq:Xi}. Note that $Q(\mathbf{0},0)=0$, which implies $\Xi(\mathbf{0},0)=0$ since $\tilde{G}(\mathbf{0},0)=0$. The constraint \eqref{eq:Xi} on $\Xi$ is $\Mb(\ab,q)\cdot\Thetab(\ab,\vv)\leq\Xi(\ab,q)$, where $\Mb$ and $\Thetab$ are defined by \eqref{eq:auxiliary}. Constructing such $\Xi$ requires additional estimates on $\Mb\cdot\Thetab$ because $\Thetab$ depends on the infinite-dimensional tail $\vv$ of the velocity's Galerkin expansion \eqref{eq:Galerkin}, whereas $\Xi$ can depend on $\vv$ only through its norm, $q=\|\vv\|$.

In previous efforts \cite{ChernyshenkoGoulart,huang2015}, the $\Mb\cdot\Thetab$ term was estimated by
\begin{equation}
	\begin{aligned}
		\Mb\cdot\Thetab&\leq|\Mb||\Thetab|\leq|\Mb|(|\Thetabab|+|\Thetabc|)\\
										&\leq|\Mb|\sqrt{2(|\Thetabab|^2+|\Thetabc|^2)},
		\label{eq:MThetaOld}
	\end{aligned}
\end{equation}
where $|\Thetabab|^2+|\Thetabc|^2=\sum_{i=1}^m(\Thetaabi^2+\Thetaci^2)$. Individual terms in the latter sum were then estimated by \cite{ChernyshenkoGoulart,huang2015}
\begin{equation}
	\begin{aligned}
		|\Thetaabi|&\leq\sqrt{\tilde{\ab}^\T\GG_i\tilde{\ab}q^2},\qquad(\GG_i)_{jk}=\scalar{\tilde{\hh}_{ij}}{\tilde{\hh}_{ik}},\\
		|\Thetaci|&\leq C_{i} q^2\,,\quad C_{i}\!=\!\|\rho(\DD_i)\|_\infty\!=\!\textstyle{\sup_{\xx}}\rho\big(\DD_i(\xx)\big),
	\end{aligned}
	\label{eq:ThetaBounds}
\end{equation}
where $\tilde{\ab}=[\begin{matrix}1&a_1&\cdots&a_m\end{matrix}]^\T$, $\GG_i$ is the $L^2$ Gram matrix of the vector fields $\{\tilde{\hh}_{ij}\}_{j=0}^m$, $\DD_i=\frac{1}{2}(\nabla\uu_i+\nabla^\T\uu_i)$ is the strain-rate tensor of the energy eigenmode $\uu_i$, and $\rho(\DD_i(\xx))$ is the spectral radius of $\DD_i(\xx)$.
Here, $\tilde{\hh}_{ij}$ is the solenoidal projection of $\hh_{ij}$ orthogonal to all the eigenmodes $\{\uu_k\}_{k=1}^m$, meaning $\scalar{\tilde{\hh}_{ij}}{\uu_k}=0$, $\nabla\cdot\tilde{\hh}_{ij}=0$, and $\tilde{\hh}_{ij}\cdot\nn=0$ at no-slip boundaries (where $\nn$ is the outward normal vector). These bounds result in an estimate $\Mb\cdot\Thetab\leq|\Mb|\sqrt{p_{\Theta}}$ with $p_{\Theta}=2\sum_{i=1}^m\big(\tilde{\ab}^\T\GG_i\tilde{\ab}q^2+C_{i}^2q^4\big)$ being a polynomial. The upper bound $|\Mb|\sqrt{p_{\Theta}}$ depends only on $(\ab,q)$ as desired, but it is not a polynomial because of the absolute value and the square root. Additional manipulations are needed to formulate purely polynomial inequalities that imply $Q<0$ (cf.\ \cite{ChernyshenkoGoulart}), which at last can be relaxed to sum-of-squares (SOS) constraints. The resulting formulation is very computationally expensive. For 2D Couette flow we have implemented it with $\ab\in\R^6$. Relative to the new approach described below, the computations were more expensive and the stability results were significantly weaker.

In the present work we derive an estimate $\Mb\cdot\Thetab\le\Xi$, where $\Xi$ is a polynomial in $(\ab,q)$. In contrast to \eqref{eq:MThetaOld}, where the Cauchy--Schwarz inequality was used immediately on $\Mb$ and $\Thetab$, we estimate
\begin{equation}
		\Mb\cdot\Thetab=\sum_{i=1}^m M_i\Theta_i\leq\sum_{i=1}^m |M_i|(|\Thetaabi|+|\Thetaci|).
\end{equation}
In general this estimate is sharper than \eqref{eq:MThetaOld}, and we still can apply the estimates \eqref{eq:ThetaBounds} derived in \cite{ChernyshenkoGoulart,huang2015}. In particular, if we can find polynomials $r_i(\ab,q)$ and $s_i(\ab,q)$ such that
\begin{align}
		|M_i||\Thetaabi|&\leq|M_i|\sqrt{\tilde{\ab}^\T\GG_i\tilde{\ab}q^2}\leq r_i,\label{eq:ridef}\\
		|M_i||\Thetaci|&\leq s_iC_iq^2,	\label{eq:sidef}
\end{align}
then $\Mb\cdot\Thetab$ is bounded above by the polynomial
\begin{equation}
	\Xi(\ab,q)=\sum_{i=1}^m\big(r_i+s_iC_iq^2\big).
	\label{eq:XiApp}
\end{equation}
The $M_i$ defined in \eqref{eq:auxiliary} depend on the Lyapunov functional $V$, so the conditions \eqref{eq:ridef}--\eqref{eq:sidef} do also. The construction of $r_i$ and $s_i$ satisfying \eqref{eq:ridef}--\eqref{eq:sidef} must be done simultaneously with the construction of a $V$ that satisfies all other constraints. For these constructions to be carried out computationally, the constraints \eqref{eq:ridef}--\eqref{eq:sidef} must be transformed into polynomial inequalities.

Condition \eqref{eq:ridef} on each $r_i$ is equivalent to the $2\times2$ positive semidefinite matrix constraint
\begin{equation}
		\begin{bmatrix}\tilde{\ab}^\T\GG_i\tilde{\ab}q^2r_i&\tilde{\ab}^\T\GG_i\tilde{\ab}q^2M_i
			\\\tilde{\ab}^\T\GG_i\tilde{\ab}q^2M_i&r_i\end{bmatrix}\succeq0\,
	\label{eq:riIneq2LMI}
\end{equation}
holding for all $(\ab,q)\in\R^{m+1}$. Multiplying this matrix on the left and right by a dummy variable $\ww\in\R^2$ gives a polynomial inequality that also is equivalent to \eqref{eq:ridef}:
\begin{equation}
	w_1^2\tilde{\ab}^\T\GG_i\tilde{\ab}q^2r_i+2w_1w_2\tilde{\ab}^\T\GG_i\tilde{\ab}q^2M_i+w_2^2r_i\geq0 \label{eq:riDummy}
\end{equation}
for all $(\ab,q,\ww)\in\R^{m+3}$. The $r_i$ cannot have any constant or linear terms because $\Xi(\mathbf{0},0)=0$, and their degree cannot exceed that of $V$.

The condition \eqref{eq:sidef} on each $s_i$ holds if $|M_i|\leq s_i$, as follows from the bound on $|\Thetaci|$ in \eqref{eq:ThetaBounds}. This can be enforced by the polynomial inequalities
\begin{equation}
-s_i\leq M_i, \qquad M_i\leq s_i.
\end{equation}
The requirement that $\Xi(\mathbf{0},0)=0$ does not restrict the lower-degree terms of the $s_i$ because $C_iq^2$ vanishes at the origin. The maximum degree of each $s_i$ can be no larger than that of $M_i$, which is two less than the degree of $V$ when using the ansatz \eqref{eq:Vansatz}.

\subsection{Final sum-of-squares conditions}

The polynomial inequalities that suffice for $V$ to be a valid Lyapunov functional, which are derived above and in the main document, can be summarized as
\begin{equation}
	\begin{aligned}
		E(\ab,q)^d+P(\ab,q)-\eps E(\ab,q)&\geq0, \\
		-\big(\tilde{G}(\ab,q)+\Xi(\ab,q)+\eps E(\ab,q)\big)&\geq0, \\
		\textstyle{\pder{V}{q^2}}&\geq0, \\
		w_1^2\tilde{\ab}^\T\GG_i\tilde{\ab}q^2r_i(\ab,q)+w_2^2r_i(\ab,q)\hspace{5mm}&{}\\
			+2w_1w_2\tilde{\ab}^\T\GG_i\tilde{\ab}q^2M_i(\ab,q)&\geq0,\\
		s_i(\ab,q)-M_i(\ab,q)&\geq0,\\
		s_i(\ab,q)+M_i(\ab,q)&\geq0
	\end{aligned}
	\label{eq:PolyLyapunov}
\end{equation} 
for $i=1,\ldots,m$. In the expressions above, $V=E^d+P$, $E=\tfrac{1}{2}(|\ab|^2+q^2)$, $P$ is a polynomial of degree no more than $2d-1$ with no constant or linear terms, $\GG_i$ and $C_i$ are constants in the estimates of \eqref{eq:ThetaBounds}, the $M_i$ are defined in \eqref{eq:auxiliary}, $\tilde G$ is defined by \eqref{eq:G}, $\Xi$ is defined by \eqref{eq:XiApp}, and $\eps>0$ is fixed. The $r_i$ are polynomials of degree at most $2d$ with no constant or linear terms, while the $s_i$ are polynomials of degree at most $2d-2$. In all polynomials, only even powers of $q$ are present. All stability computations reported here used $d=2$, corresponding to quartic Lyapunov functionals $V$, cubic $P$, quartic $r_i$, and quadratic $s_i$. In all cases we fixed $\eps=2\cdot10^{-5}$.

As noted in the main document, verifying the nonnegativity conditions in \eqref{eq:PolyLyapunov} is computationally intractable (NP-hard \cite{Murty1987}). Therefore, we strengthen them to more tractable SOS conditions. To search for a Lyapunov functional $V$, the tunable variables in \eqref{eq:PolyLyapunov} are the coefficients of the polynomials $P$, $r_i$, and $s_i$. Every expression that must be SOS is linear in these polynomials, which is essential. Polynomial optimization problems subject to SOS constraints can be converted into semidefinite programs only if the SOS expressions are linear in the tunable parameters. Indeed, there are many possible ways of deriving an upper bound $\Mb\cdot\Thetab\le\Xi$, but only certain estimates lead to a formulation in which the tunable parameters appear linearly. The estimates used here have this property, as do the ones derived in \cite{ChernyshenkoGoulart}. Thus, having replaced each inequality in \eqref{eq:PolyLyapunov} with an SOS condition, we used the software YALMIP \cite{Lofberg2004,Lofberg2009} to formulate an equivalent semidefinite program and the software MOSEK v8.0.0.80 \cite{mosek} to solve it numerically.

To decrease computational cost and improve numerical conditioning, symmetries of the governing dynamics \eqref{eq:auxiliary} can be used to anticipate symmetries of the polynomials $P$, $r_i$, and $s_i$, whose ans\"atze then can be chosen to impose these symmetries. The symmetries of the truncated system $\tder{\ab}{t}=\ff$ depend on the choice of eigenmodes. With the 6-mode set defined in the main document, for instance, the system is invariant under the rotation $(a_1,a_2,a_3,a_4,a_5,a_6)\to(a_1,a_2,a_4,-a_3,a_6,-a_5)$. Thus, our ans\"atze for $P$, $r_i$, and $s_i$ omitted all monomials that are not invariant under the transformations generated by this rotation.
To ease the detection of symmetries and increase sparsity of the tensors in \eqref{eq:auxiliary}, our computations for 2D plane Couette flow used the spatial domain $(0,L)\times(-\tfrac{1}{2},\tfrac{1}{2})$, rather than $(0,L)\times(0,1)$.

\subsection{Numerical solution of the energy eigenproblem}

To solve the energy eigenproblem \eqref{eq:eigen} in the case of 2D plane Couette flow, we define a stream function $\psi$ such that $\uu=(\pder{\psi}{y},-\pder{\psi}{x})$. Then \eqref{eq:eigen} implies
\beq
\lambda\Delta\psi=\psderm{\psi}{x}{y}+\frac{1}{\Rey}\Delta^2\psi
\label{eq:eigen psi}
\eeq
with the no-slip boundary conditions $\nabla\psi(x,\pm\frac{1}{2})=\mathbf{0}$. Fourier decomposition in $x$ gives an equation for $\hat\psi(\alpha,y)$, where the streamwise wavenumber $\alpha$ is an integer multiple of $\frac{2\pi}{L}$. The solution is $\hat{\psi}(\alpha,y)=\sum_{j=1}^4c_je^{i\beta_jy}$, where the $\beta_j$ are roots of the characteristic polynomial
\beq
p_\psi(\lambda,\Rey,\alpha,\beta)=\textstyle{\frac{1}{\Rey}}(\alpha^2+\beta^2)^2+\lambda(\alpha^2+\beta^2)-\alpha\beta.
\label{eq:CharPoly}
\eeq
(The $\hat\psi$ ansatz changes slightly when there are repeated roots, but this occurs only on a parameter set of measure zero.) The no-slip boundary conditions at $y=\pm\tfrac{1}{2}$ imply $\BB\CC=\mathbf{0}$, where $\CC=[\begin{matrix}c_1&c_2&c_3&c_4\end{matrix}]^\T\in\mathbb C^4$ and
\beq
	\BB=\begin{bmatrix}\cos(\frac{1}{2}\beta_1)&\cdots&\cos(\frac{1}{2}\beta_4)\\[1pt]
		\sin(\frac{1}{2}\beta_1)&\cdots&\sin(\frac{1}{2}\beta_4)\\[1pt]
		\beta_1\cos(\frac{1}{2}\beta_1)&\cdots&\beta_4\cos(\frac{1}{2}\beta_4)\\[1pt]
		\beta_1\sin(\frac{1}{2}\beta_1)&\cdots&\beta_4\sin(\frac{1}{2}\beta_4)
	\end{bmatrix}.
\eeq
The matrix $\BB$ depends on $(\lambda,\Rey,\alpha)$ through the roots $\beta_j$ of the characteristic polynomial \eqref{eq:CharPoly}. For given $\Rey$ and $\alpha$, eigenvalues $\lambda$ of \eqref{eq:eigen psi} can be found from the fact that $\BB$ must have a nontrivial kernel. In such cases, $\lambda$ is an eigenvalue of the energy eigenproblem \eqref{eq:eigen psi} at the corresponding $(\Rey,\alpha)$ values, and the null vector $\CC$ can be used to construct $\hat{\psi}(\alpha,y)$. The velocity field of the eigenmode is $\uu=\big(\pder{\psi}{y},-\pder{\psi}{x}\big)$, where $\psi(x,y)=2\Real(\hat{\psi}(\alpha,y)e^{\I\alpha x})$. Each nonzero-$\alpha$ eigenmode represented by $\CC$ is $L^2$-orthogonal to the quarter-phase shifted eigenmode represented by $e^{\I\pi/2}\CC=\I\CC$. We take both of these modes to span the $\lambda$-eigenspace. All eigenmodes $\uu$ are normalized by their $L^2$ norms, so the resulting set of modes $\{\uu_1,\ldots,\uu_m\}$ is $L^2$-orthonormal.

At each fixed $(\Rey,\alpha)$, we used a nonlinear root solver to find $\lambda$ at which $\BB$ is singular, corresponding to $\lambda$ that are eigenvalues of \eqref{eq:eigen psi}. Our implementation searched for $\lambda$ that locally minimized the smallest singular value of $\BB$, as computed by a singular value decomposition. This numerical approach was more robust than searching for zeros of $\det(\BB)$ or tracking individual eigenvalues of $\BB$. Only a certain number branches were tracked at each multiple of $\frac{2\pi}{L}$. In the special case $\alpha=0$, the streamwise-independent eigenfunctions can be found analytically starting from \eqref{eq:eigen}. The eigenvalues are $\lambda=-\tfrac{1}{\Rey}(k\pi)^2$ for $k\in\N$, with eigenfunctions $\uu=\tfrac{2}{L}\big(\cos(k\pi y),0\big)$ when $k$ is odd and $\uu=\tfrac{2}{L}\big(\sin(k\pi y),0\big)$ when $k$ is even.

Once a set of eigenmodes $\{\uu_1,\ldots,\uu_m\}$ was selected, $\ff$ and the $\hh_{ij}$ were computed according to the formulas in \eqref{eq:auxiliary}. The largest eigenvalue not associated to these modes gives the value of the constant $\kappa$. Because we use the estimate \eqref{eq:ThetaBounds} to bound $\Mb\cdot\Thetab$, the $\GG_i$ and $C_i$ appearing in \eqref{eq:ThetaBounds} had to be computed as well.

To compute the matrices $\GG_i$, we first need the $\tilde\hh_{ij}$, which are the solenoidal projections of the $\hh_{ij}$ onto the orthogonal complement of the selected eigenmodes. First, the solenoidal projection $\hh_{ij}^{\mathrm{div}}$ was calculated by solving a Poisson problem of the form $\Delta\phi=\nabla\cdot\hh_{ij}$ with boundary conditions $\nabla\phi\cdot\nn=\hh_{ij}\cdot\nn$, in which case $\hh_{ij}^{\mathrm{div}}=\hh_{ij}-\nabla\phi$ with $\hh_{ij}^{\mathrm{div}}\cdot\nn=\mathbf{0}$ wherever $\uu$ has no-slip boundary conditions. This computation was done in Fourier space by exploiting the Fourier decomposition of the initial $\hh_{ij}$. Then, the projections of $\hh_{ij}^{\mathrm{div}}$ onto $\uu_1,\ldots,\uu_m$ were subtracted from $\hh_{ij}^{\mathrm{div}}$ to produce the $\tilde{\hh}_{ij}$. Lastly, the $\GG_i$ were the $L^2$ Gram matrices associated to $\{\tilde{\hh}_{ij}\}_{j=0}^m$, as defined in \eqref{eq:ThetaBounds}.

The eigenvalues of $\DD_i(\xx)$ appearing in the estimates \eqref{eq:ThetaBounds} were computed explicitly. Incompressibility of the flow implies $\mathrm{tr}(\DD_i(\xx))=0$ for all $\xx$, and in the two-dimensional case this means that $\DD_i(\xx)$ has two opposite eigenvalues whose magnitude is the spectral radius. Hence the spectral radius is 
\begin{equation}
	\rho(\DD_i(\xx))=\sqrt{(\textstyle{\psderm{\psi_i}{x}{y}})^2+\tfrac{1}{4}(\Delta\psi_i)^2},
\end{equation}
where $\psi_i$ is the stream function of the energy eigenmode $\uu_i$. The maxima of the $\rho(\DD_i(\xx))$ over all $\xx$ in the domain were found numerically using constrained optimization methods, giving the values $C_i=\|\rho(\DD_i)\|_\infty$.

\subsection{Independent check of Lyapunov functional}

\setcounter{figure}{2}
\begin{figure}[h]
\vspace{-8mm}
\centering 
\includegraphics[width=6.65cm]{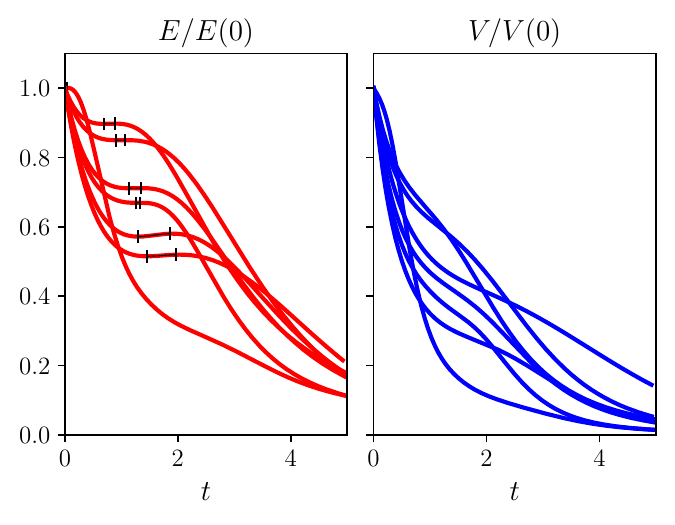} 
\vspace{-5mm}
\caption{Transient evolution of the perturbation energy $E$ and the quartic Lyapunov functional $V$ computed via SOS methods (using the 13-mode system described in Fig.\ \ref{fig:Eigenmodes}) for the parameters $(\Rey,L)=(240,2)$. The 7 trajectories displayed correspond to the only cases out of $10^4$ random initial conditions simulated that lead to transient growth of $E$ (with time-intervals of growth indicated by vertical lines).\label{fig:RandomTraj}\vspace{-2mm}}
\end{figure}

To independently check that the $V$ constructed by our SOS methods are valid Lyapunov functionals that establish global stability of 2D plane Couette flow, we numerically integrated the Navier--Stokes equations \eqref{eq:NS}--\eqref{eq:inc} from many different randomly generated initial conditions. At each time step, values of $V$ and the energy $E$ were computed by evaluating spatial integrals. More precisely, $E(\uu)=\frac{1}{2}\int |\uu|^2\dx$ is a single spatial integral, whereas $V(\uu)=E^2+P(\ab,q)$ with $P(\ab,q)$ being a cubic polynomial coming from an SOS computation, is dependent on the spatial integrals $a_i(\uu)=\int\uu\cdot\uu_i\,\dx$ and $q=(2E-\ab\cdot\ab)^{1/2}$.

Direct numerical simulations of \eqref{eq:NS}--\eqref{eq:inc} were carried out with the software \texttt{dedalus} \cite{dedalus} using a spectral spatial discretization with 64 Fourier modes in the $x$-direction and $32$ Chebyshev modes in the $y$-direction. A two-stage Runge--Kutta method was used for time-stepping.

The velocity field of each initial condition was derived from a randomly generated stream function. To generate the stream function, the coefficients of its spectral modes were chosen randomly. To ensure smoothness of the initial condition, the random coefficients were required to decay exponentially beyond a randomly selected modest number of cutoff modes in $x$ and $y$. The resulting stream function was multiplied by $(\tfrac{1}{4}-y^2)^2$ in order to satisfy no-slip boundary conditions.

As an example we selected $(\Rey,L)=(240,2)$ as parameters, where 2D plane Couette flow is energy-unstable---modes $(1,1)$ and $(2,1)$ in Fig.\ \ref{fig:Eigenmodes} produce initial energy growth---whereas our SOS computations produce a quartic Lyapunov functional $V$ depending on the the 13 modes described in the caption of Fig.\ \ref{fig:Eigenmodes}. According to the analysis in the main text, this $V$ must decrease monotonically in time for any solution of \eqref{eq:NS}--\eqref{eq:inc}. We carried out simulations starting from $10^4$ random initial conditions. Only 7 trajectories displayed transient energy growth but, as expected, $V$ decreased monotonically in time for all $10^4$ trajectories. Figure \ref{fig:RandomTraj} shows the time evolution of $E$ and $V$ for the 7 trajectories with transient energy growth.

%\end{document}

\end{document}